# Secure Extensibility for System State Extraction via Plugin Sandboxing

*Sahil Suneja and Canturk Isci, IBM Research*

## Abstract

In this paper, we introduce a new mechanism to securely extend systems data collection software with potentially untrusted third-party code. Unlike existing tools which run extension modules or plugins directly inside the monitored endpoint (the guest), we run plugins inside a specially crafted sandbox, so as to protect the guest as well as the software core. To get the right mix of accessibility and constraints required for systems data extraction, we create our sandbox by combining multiple features exported by an unmodified kernel. We have tested its applicability by successfully sandboxing plugins of an opensourced data collection software for containerized guest systems. We have also verified its security posture in terms of successful containment of several exploits, which would have otherwise directly impacted a guest, if shipped inside third-party plugins.

## 1 Introduction

Systems data collection is an essential component in every cloud monitoring setup [108]. There exist several systems monitoring and data collection tools, like Collectd [39] and Nagios [16], which extract system-level state, such as the inventory of all running applications, open connections, etc. Such tools typically follow an extensible model, presumably to gain widespread adoption, where code from users or third-party developers can be incorporated to run with the core software, in the form of modules, plugins, classes, etc. (collectively referred to as 'plugins' from here onwards). These plugins implement state-specific collection logic to extract the relevant system state, for example, querying the procfs to gather process-level metrics, or querying the package database to gather names and versions of installed packages. Figure 1(a) shows a typical data collection setup in container clouds. Here, the core monitoring software, running on a host, commands and controls different data collection plugins. As shown in the figure, these plugins may be host-resident, or running on the monitored endpoints, called *guests*.

Since these plugins usually interface directly with the guest and the host, the plugins have the ability to impact their operations. The plugins may be buggy or malicious, causing data corruption or resource hogging, for example. There exist official CVE entries [1] for vulnerabilities in plugins leaking sensitive information (such as application login credentials) [7–9], enabling arbitrary command execution [2,4,6], causing Denial of Service (DoS) [1,5,11], amongst other security concerns.

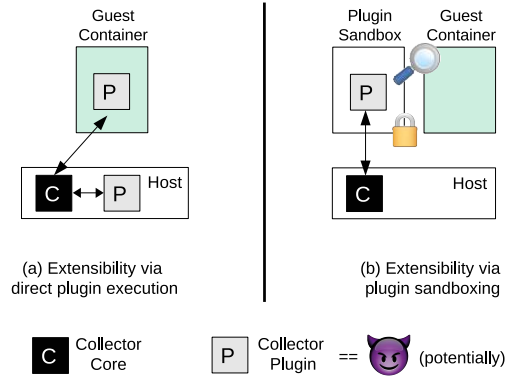

Figure 1: Systems data collection methodology in a container cloud. The monitoring software core (C) interfaces with several plugins (P). (a) Existing approaches: Direct plugin execution at host as well as inside monitored guest. (b) Our approach: Sandboxed plugin execution outside the guest's context.

In case of the popular Nagios montoring software, plugin vulnerabilities account for 23% of all CVEs associated with Nagios[2]. Many of the remainder vulnerabilities in the software core are further exploitable by potentially malicious plugins, by attacking the software core's otherwise legitimate interface to communicate with external entities, such as via crafted payloads [10, 43].

Clearly, opening a tool to interface with third-party plugins increases its vulnerability exposure, which needs extra guarding beyond the tool developer's responsibility of securing the core software. In order to understand how existing tools deal with this, we performed a survey of the security posture of their extensibility models. We observed that a few tools do incorporate some security emphasis in their design, such as encryption, authentication and authorization [38, 70, 74]. While a few others publish certain security-oriented guidelines for plugin authors to follow, and end users to ensure, such as non-root operations, directory permission lock-down, input validation, and source code review [85, 86, 88, 101, 107]. But, *most tools distance themselves from a direct impact by transferring the security onus onto the guest*. A common trend is to follow a client-server separation model [86, 95, 105], where the host running the software core is protected from third-party plugins running inside the guest.

This guests' exposure to third-party plugins means that the guest still remains susceptible to malicious or vulnerable plugins (such as the aforementioned CVEs). There is thus a need for a better solution, one that protects both–the monitoring software core running on the host, *as well as* the moni-tored guest–from the plugins' side-effects. The challenge in securely running untrusted system-state-extraction plugins, lies in balancing the levels of accessibility and constraints afforded to them. One one hand, we need to properly sandbox untrusted third-party plugin code, so that it can't harm either the guest or the host, either directly (e.g., subverting processes, or leaking secrets), or indirectly (e.g. DoS, or acting as botnet). And, from an accessibility perspective, we need to provide these plugins *secure* access to *privileged* resources *outside* the walls of a typical sandbox, specifically–the target guest's memory and disk state.

---

[1] Common Vulnerabilities and Exposures database for publicly-known cybersecurity vulnerabilities. https://cve.mitre.org/

[2] Stats for the opensourced Nagios-core software only; not the Nagios XI enterprise product. From https://www.cvedetails.com/

In this paper, we present a new technique to safely run potentially adversarial third-party plugins, with systems monitoring and data collection software. Our approach to enable secure extensibility is to run extension modules or plugins outside the guest context, at the host, to protect the guest and avoid any guest modifications. And to protect the host as well as the software core, we isolate these plugins in a specially crafted sandbox, as shown in Figure 1(b), The sandbox isolation, as well as secure access to the guest's state, is achieved by combining multiple constructs of an *unmodified* kernel. We follow the principle of *least privilege* [102] in our sandbox design, starting with running the plugins as unprivileged entities, then giving them access rights via *namespaces* and *capabilities*, and finally restricting their impact potential via *seccomp* and *netfilter*. In addition to these five constructs for access controls, we also use *cgroups* to enforce resource constraints on the plugin sandbox.

We highlight our sandbox' strong security posture by running exploits across 10 different attack vectors inside it, and verifying it's ability to contain all exploits, which would have otherwise directly impacted the guest (or the host), had they been shipped inside third-party plugins. We demonstrate our sandbox' applicability by using it to isolate plugins of an existing container monitoring software, with most plugins running unmodified, while the rest requiring only minor modifications. We also measure the overhead our sandbox introduces to a normal systems data collection flow, in terms of sandbox creation latency (344ms), running time degradation (an increase from 26ms to 30ms per monitoring cycle), and resource consumption (zero overhead).

While we focus on containers as our target runtime, the sandboxing technique employed is equally applicable for securely extending in-VM or host-local data collection software. One difference from the container guest targets is that the plugins would continue to operate inside the VM (or the host), but isolated in the same sandbox as presented in this paper, while accessing the VM's (or the host's) state securely.

This paper makes the following contributions:

- A survey of the extensibility and security models of popular system state extraction tools.

- A read-only permissive, yet isolated sandbox, created by combining constructs from an unmodified Linux kernel.

- An sidecar-container-based plugin sandbox prototype implementation, which we have opensourced.

The rest of the paper is organized as follows. We present a survey of existing systems data collection tools in Section 2. We clarify our threat model in Section 3, the design of our sandbox in Section 4, and a prototype implementation using sidecar containers in Section 5. In Section 6, we evaluate the security, applicability and overhead of our sandbox, and compare it with existing approaches in Section 7.

## 2 Security Posture of Existing Tools

In the context of cloud monitoring, we performed a survey of 17 popular tools which can be employed for container data collection, to understand the security posture of their extensibility offerings. Table 1 compares them in terms of a tool's extensibility model, the associated security concerns which may arise while running third-party code, as well as any security emphasis in the tool's design to protect against untrusted code. The table cells mention whether and how a property is applicable to a tool (as a 'Y' (yes), or a 'N' (no)).

The way in which a tool supports extensibility is mentioned in the corresponding cell of column 2. Different tools use different terminology for their extensibility mechanism- plugins, classes, packages, integrations, exporters, instrumentation libraries, etc. No security concerns are flagged ('N' for cells in columns 4 and 5) for tools which either (i) do not provide any, or only a limited (alpha/preview only) support for extensibility (e.g., Aqua, Cadvisor, Heapster), or (ii) support extensibility not for data collection, but other higher level rules and security policies, such as network isolation rules, security checklists, etc. (e.g., Twistlock, Neuvector), or (iii) do not interface directly with the target endpoint (e.g., Anchore and Clair (image copy scanning), Tenable (port scanning)).

In terms of safeguards against third-party plugins, we can see in column 3 that a few tools, such as Collectd and Sensu, do incorporate some security emphasis in their design, such as encryption, authentication and authorization [38, 70, 74]. Prometheus publishes best practices for writing exporters, instrumentation, and labeling for the data collection output flowing from the guest to the host [94]. Others, such as Nagios, Tenable, and New Relic, publish some security-oriented guidelines for plugin authors to follow, and end users to ensure—such as ensuring non-root operations, directory permission lock-down, input validation, and source code review [85, 88, 101, 107]

Despite the varying degree of security emphasis, potential security concerns exist on the guest and/or the host side (a 'Y' in the column 4/5 cells) depending upon the extensible tool's design, in terms of the environment (in-guest or on-host) in which the plugins are supposed to run. An impact may be direct- when plugins run as root, or indirect- such as a fork bomb, botnet, DoS kind of behaviour. Tool-specific reasoning behind any security concerns are mentioned in the corresponding cell.

| Properties / Tools | Extensible? | Security emphasis to protect against 3rd party components? | Security concerns on host when running 3rd party components? | Security concerns on guest when running 3rd party components? |
|---|---|---|---|---|
| Aqua | N | N | N (no extensibility) | N |
| Twistlock | N (different meaning; not in terms of data collection) | N | N (no extensibility) | N |
| Neuvector | N (different meaning; not in terms of data collection) | N | N (no extensibility) | N |
| Cadvisor | N (limited/alpha) | N | N (no extensibility) | N |
| Heapster | N (limited/alpha) | N | N (no extensibility) | N |
| Anchore | Y (scripts) | N | N (can run on separate host) | N (runs against separate instance) |
| Clair | Y (package/class) | N | N (can run on separate host) | N (runs against separate instance) |
| Splunk | Y ('add-ons') | N | N (remote execution) | Y |
| Prometheus | Y (instrumentation libs and 'exporters') | Y (best practices guidelines) | N (remote execution; narrow interface) | Y |
| Sysdig | Y (scripts) | N | Y (scripts run as root) | Y (depends on host health) |
| Datadog (dd) | Y (statsd and 'integrations') | N | Y (if dd-agent runs in host; integrations == run scripts locally) | Y (if dd-agent runs in guest) |
| Collectd | Y (plugins) | Y (selinux, encryption, signing) | Y (fragile interface; documented CVEs) | Y (plugins run as root) |
| Agentless Crawler- ASC | Y (plugins) | Y (code review, CI/CD testing) | Y (when plugins [default = root] run on host) | Y |
| Nagios | Y (plugins; external commands) | Y (code review, security guidelines) | Y (accepts direct commands; documented CVEs) | Y |
| New Relic | Y (plugins) | Y (security guidelines) | N (remote execution) | Y |
| Sensu | Y (plugins/'checks') | Y (run only server-provided (subscription mode) / local-defined (safe mode) checks | N (remote execution) | Y (checks may run as root [default = non-root]; + indirect impact) |
| Tenable.io | Y (plugins) | Y (signed plugins) | N (can run on separate host) | N (no in-guest operation) |

Table 1: Security posture of existing tools w.r.t. extensibility; Guest remains vulnerable ('Y' in last column) in most extensible tools.

Overall, we observe that a common approach amongst the extensible tools is to follow a client-server separation model [86, 95, 105], where the third-party plugin code is run inside the monitored guest, while the software core runs on a separate host environment. Beyond basic metrics gathering, most solutions run data collection logic on the guest. Although this may protect the host running the software core from the untrusted plugins' impact, the security onus gets transferred to the guest. This can be observed, for example, for Prometheus, Sensu, New Relic, and Splunk, where the host is spared the plugin execution impact, but not the guest (an 'N' in column 4, and a 'Y' in column 5). On the other hand, the host remains vulnerable to a direct impact by potentially malicious third-party code, for tools which do not enforce this separation and allow running plugins on the host itself. This includes tools such as Collectd, Nagios, Datadog, as well as the 'agentless' class of monitoring solutions [91], such as the Agentless System Crawler (ASC).

Our sandbox enables securely running the untrusted plugins while protecting both the guest and the host.

## 3 Threat Model

We focus on the general setting of a systems data extraction software running in a container cloud. The goal is to restrict the plugin's operational environment enough so that it can't do anything malicious. This includes direct impact such as process corruption, as well as indirect interference such as botnet or fork-bomb behaviour. For the purpose of systems data collection, we are okay with giving the plugin full read-only (R/O) visibility into a guest container's state, including possibly any secret keys in there. Multiple instances of the same plugin may exist at the same time. The adversary can author multiple plugins which can operate against the same guest container. The adversary can also run their container (as a regular unprivileged cloud tenant) on the same host as target guest container. We don't focus on protecting plugins from each other (they can be put in separate sandboxes if need be).

### 3.1 Assumptions

A. We trust the host admin and also assume the host itself isn't subverted.

B. Although a plugin may ship sensitive data, masked as a valid system-state feature (ex. CPU metrics, or file hash), to the monitoring backend, we assume gates are in place at the backend to allow access to extracted data only to the legitimate user (e.g. the container owner).

C. We also assume certain checks are in place to prevent a plugin from colluding and sharing sensitive information, with another plugin instance or a higher capability process, via host-level side channels (locks, caches, etc.) [66, 116].

D. We also assume certain properties about the host kernel's implementation: (i) reading the `procfs` has no impact on a process, (ii) different kernel features play well with each other: i.e. capabilities, seccomp, and iptables are properly namespaced, and feature combinations, such as capabilities + seccomp, are valid and do not weaken or break an abstraction or introduce a new security hole. To this end, we performed a detailed exploration of the kernel source code,to verify the containment of any potential exposure which capabilities may introduce. We also ran the relevant exploits and confirmed their failure to cause a break-out (Section 6.1).

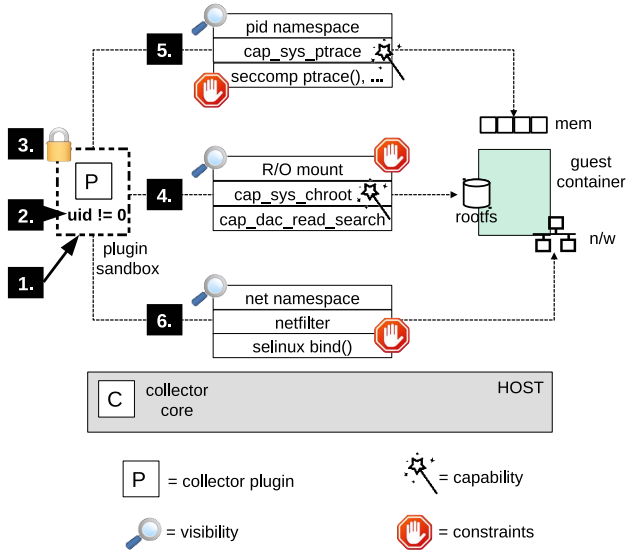

Figure 2: Plugin sandbox design. The monitoring software core (C) running on the host, interfaces with the data collection plugin (P) running outside the guest context in a sandbox. The numbered labels represent the sequence of operations performed to isolate the plugin via namespacing (1), de-privileging (2) and cgroup-ing (3), as well as to enable secure access to the guest's disk, memory, and network state (Steps 4-6) via a blend of visibility, capability, and constraints.

## 4  Design

The challenge in securely running untrusted system-state-extraction plugins, lies in balancing the levels of accessibility and constraints afforded to the them. In terms of specific restrictions, we want to prevent the plugins from: (i) impacting the guest's and the host's execution: this includes direct impact such as process subversion, as well as indirect interference such as DoS via resource hogging, (ii) communicating with the outside world: leaking sensitive information or acting as a botnet, and (iii) leaking information to host-local accomplices. And, from an accessibility perspective, we need to provide these plugins *secure* access to *privileged* resources *outside* the walls of a typical sandbox, specifically–the target guest's memory and disk state.

To get this desired accessibility-constraint blend in our sandbox, we tap into the kernel to pick relevant constructs and fine tune them. As a brief description of the sandbox design, we use *namespaces*-based view abstraction [19] to isolate the plugins away into a separate environment than the target guest's. Selective access to the target guest's environment is provided to the plugins via namespace sharing [21]. The plugins are run as *unprivileged* entities, with read-only access to privileged guest state enabled via *capabilities*-based selective power conferral [18]. Finally, the plugins' impact potential is restricted by using *seccomp*-based syscall filter-ing [20], *netfilter*-based network packet filtering [17], and *cgroups*-based resource limits [3].

Figure 2 shows the overall design of our sandbox in a containerized environment, numbered to reflect the sequence of operations performed to isolate the plugin (Steps 1-3), as well as to enable secure access to guest's state (Steps 4-6). We detail the sandbox build operations below.

**1. View Isolation.** First, the plugin is put into its own set of namespaces (aka the *sandbox*), separate from the the guest and host. It can thus only view its private set of processes, mount-points, network devices, user/group IDs, and inter-process communication objects. This isolates the plugin away, and prevents it from having any communication with entities outside its sandbox. This includes communication via the filesystem, the network, or through inter-process communication mechanisms such as shared memory or message queues.

**2. De-privileging.** Next, following the principle of least privilege [102], the plugin is made an unprivileged entity, by mapping its user ID inside its sandbox' namespace to a non-root user ID on the host. This takes away a substantial amount of power from the plugins to alter system state (which it is awarded access to, starting from Step 4 onwards).

**3. Resource Isolation.** The sandbox is then put in a separate cgroup for resource isolation. Limits can thus be enforced on the plugin's process count, CPU, memory and disk usage, preventing it from indirectly impacting the guest and the host.

*Now, since the plugin is in its own set of namespaces and cgroup'ed, it can not harm the guest or the host, under the assumptions of Section 3.1 However, this also means that it can not 'see' the target guest state, which it needs access to for carrying out its data extraction task. Since data-collection plugins essentially extract state residing in memory or disk, we thus need to give the sandboxed plugin secure read-only access to guest state, as shown in steps 4-6 in Figure 3*

**4. Access to disk state.** Isolated in its own mount namespace, the plugin cannot see the guest's disk-level system state such as configuration files, logs, package databases, etc. Thus, to be able to access this state, the guest container's root filesystem (rootfs) is mounted read-only inside the sandbox.

But, since the plugin's and the guest's user IDs are different, the plugin may be unable to read the guest's files in the mounted rootfs, because of discretionary-access-control (DAC) permission checks. The plugin sandbox is thus granted CAP_DAC_READ_SEARCH capability to see guest container's files (only reads, no writes).

Now, while the plugin expects to find the relevant files at paths relative to the root directory ('/'), the actual state exists in the mounted guest rootfs (say at '/some/location/'). Thus, the execution environment should be set up to point to the correct root directory, i.e. chroot('/some/location'), so that the plugin and any imported libraries can work as-is, believing they are operating on the guest's '/'. But, since an unprivileged user (and thus the plugin) can't call the chroot syscall, CAP_ SYS_CHROOT capability is also granted to the sandbox to enable this view change during data collection.

**5. Access to memory state.** Similar to the mount namespace restriction, since the plugin is in its separate PID namespace, it cannot see the guest's process state. The guest container's PID namespace is thus shared with the plugin sandbox, giving it access to the guest's memory state via `procfs`.

However, despite the read-permission-check exception granted in Step 4, the plugin can still not see the files or sockets opened by the guest processes. Basically, reading a process' open files or sockets, by dereferencing `/proc/<pid>/fd/*` symlinks, requires that the ptrace access mode PTRACE_MODE_READ (less powerful, for read-only operations) be set, amongst other flags. But there is no direct mechanism to grant just these credentials to a userspace entity. Thus, as the nearest alternative, the more-powerful CAP_SYS_PTRACE capability is granted to the unprivileged plugin, in order to read such memory state. This is indeed an overkill and an artifact of the coarse grained capabilities in the Linux kernel. (vs. Capsicum's [115] fine-grain capabilities).

Since the ptrace capability is too powerful (giving the plugin an ability to kill/hang/corrupt guest processes), we use seccomp to block the 'harmful' system calls which this capability enables–`ptrace()` and `process_vm_writev()`.

Thus, by blending the ptrace capability with seccomp, we achieve our goal of giving the plugin just enough power to only read, and not impact, the guest's memory state.

**6. Access to network state.** As before, being in its own network namespace prevents the plugin from seeing the guest's network connections. To enable reading such state, the guest container's network namespace is shared with the sandbox.

The plugin can still not perform any packet-level data collection because of its unprivileged status. For passive collection, such as netflow data [35], CAP_NET_RAW capability can also be granted to the the sandbox. This does not allow in-line packet modification and insertion.

However, access to the guest's network namespace does open up two avenues of nefarious actions by the plugin. First, although the unprivileged plugin cannot disrupt the guest's network connections, it can use it to communicate with the outside world– create backdoors, steal secrets, act as botnet, etc. To avoid this, netfilter-based packet filtering is employed to block a plugin's access to the outside world, except for possibly a secure communication channel to ship out collected data to the monitoring backend.

The second concern is a potential DoS attack by the plugin, where the plugin can hoard all of the unprivileged network ports the guest has access to (cgroups doesn't prevent this) This can then prevent a guest application to communicate via network, if it hasn't already bound to its desired port. One option can be to use seccomp to block the `bind()` syscall by the plugin. But a plugin may legitimately be using local ports for data collection, or binding to a unix domain socket [109][3]. Alternatively, we use SElinux to allow only a few ports for the plugin to bind to.

**7. Access to resource stats.** Since the plugin is run in a separate cgroup for resource isolation, in order to gather the guest container's resource usage stats, the plugin is also granted access to the guest's cgroup filesystem. DAC settings ensure read-only behaviour by default.

**8. Access by software core.** To allow the software core to command and control the isolated plugin's execution, a secure communication path is set up between the plugin sandbox and the core software. Section 5.1 describes a potential scenario. This is the only way a plugin communicates with an external entity- only options being the software core and the monitoring backend (not shown in the figure).

### 4.1 Limitations

**Artifacts.** Since the plugin sandbox shares the guest container's PID namespace, the latter can 'see' processes from a foreign user seemingly running inside its context. Although it does 'pollute' the guest's view, but this is harmless because of the checks we put in place to avoid any guest impact. This can be 'fixed' by modifying relevant utilities, such as `ps`, or `docker top`, to mask the plugin container's processes.

**Visibility vs. Security Trade-off.** In our sandbox, we also block localhost (127.0.0.1; namespaced) access. This is because, even though we have guards in place to prevent exploits to the guest container via memory (privilege separation, no `ptrace()`, R/O `/proc/<pid>/mem`, no `/dev/mem` access), or disk (R/O rootfs), a plugin may abuse its proximity to the guest container to mount network attacks over the localhost interface. For example, not all ports of the guest container may be exposed to the outside world, some may just be internal, accessible over a VPN. In this case, the host-local plugin sandbox provides a better avenue for attack against a vulnerable application running inside the guest container (Section 6.1). Although blocking the access to localhost thwarts these attacks, but it may disable a legitimate plugin which collects runtime state by accessing local services (e.g. querying number of active workers from apache's status webpage). Although app-specific solutions can be employed (like allowing http/GET over localhost, but not POST, for the apache case), but it is not a generic fix across all applications. There thus exists a visibility vs. security trade-off.

**Active-events vs. Isolation Trade-off.** The sandbox design allows event-based data collection, e.g. (i) perf-event sampling such as via `perf_event_open()` (allowed by CAP_SYS_PTRACE capability), or (ii) network packet capture (allowed when CAP_NET_RAW is conferred). However, these may negatively impact the guest's performance (and can thus optionally be disallowed). Syscall tracing, on the other hand, cannot be performed inside the sandbox, due to a disallowed `ptrace()` for guest's protection (Section 4, Step 5). There thus exists a trade-off between enabling active-events collection vs. isolating its impact, although protecting an entity from event-sampling overhead is an orthogonal problem.

---

[3]Seccomp-BPF won't solve the selective bind() case since a filter won't be able to dereference a pointer– the syscall arg containing the port info.

## 5 Implementation

There exist several container engines for a container's life-cycle management, such as LXC, Docker, rkt and OpenVZ amongst others. We use Docker containers [12] in our implementation, but our sandboxing approach, being based upon fundamental kernel-level constructs, is applicable to the other engines as well. We did not need (or want) to modify the kernel; we were able to build our sandbox with already-exported kernel functionality. Furthermore, we did not have to modify Docker either, since it already exposes various runtime flags which allowed us to invoke the relevant OS-level constructs to implement our desired sandbox, as described in Section 4. *Note that a regular container by itself, including a Docker container, is insufficient to serve as a data collection sandbox.*

We use a sidecar container pattern [15] to implement our sandbox. Specifically, the plugins are run inside a separate container than the target guest container. This puts the plugin container in its own private set of namespaces, except for PID and NET namespaces, which are shared with the guest container using commandline options `-pid, -net`. Being a separate container, the plugins are put into a separate cgroup by default. The corresponding limits on the container's resource usage can be set via `-cpus, -memory, -pids-limit`, etc. The plugin container is run as an unprivileged user, different than the guest container user, in a separate user namespace than the host's[4]. The guest container's `rootfs` (in read-only mode), as well as cgroup filesystem, is also mounted inside the plugin container by using the `-volume` flag. All relevant capabilities are provided to the plugin using Linux's `setcap` utility, while also enabling the same in the plugin container's bounding set via `-cap-add` commandline option.

In terms of restraints, the relevant seccomp rules are added to the plugin container via `-security-opt seccomp` flag. Relevant iptable rules are set up within the plugin container's network namespace context using `ip-netns` utility. Cgroups-based matching (via the `net_cls` controller) is used to drop packets to/from the plugin container. In our prototype, we use the plugin container's `rootfs` as a communication channel.

### 5.1 End-to-end Execution Environment

In this Section, we describe a potential end-to-end scenario of running a plugin, inside a sandbox, against a guest container.

**Where do the plugins come from?** The plugins can be third-party or guest-authored (no need to sandbox then!). They may be hosted at any third-party server, or a central repo (like Nagios Exchange [87]), or in an object store under the guest user's account. In a particular file in the guest container's `rootfs` (say *plugins-to-run*), the guest user specifies weblinks to download the plugin files from, or simply the plugin name/ID, depending upon the hosting scenario.

---

[4] Docker currently does not support per-container user namespace [83]

**How do the plugins get run?** One option is for the software core on the host to fetch the plugins mentioned in the *plugins-to-run* file, put them inside a plugin container, and initiate, say, a *plugin-runner* process in there, with relevant arguments e.g. collection frequency. Alternatively, the software core creates a plugin container first, then initiates the *plugin-runner* process in there, which further fetches the plugins from the hosting store, as listed in the *plugins-to-run* file (accessible via R/O mounted guest `rootfs`). The difference in case of the second option is that the network-blocking needs to be set up *after* the plugin fetch step. The *plugin-runner* then runs the plugins at the set frequency. *Note that the since the plugins never get run outside the sandbox, an adversarial plugin's attempts to sense its environment before exposing its attack is futile.*

**How is the plugin output collected and sent to the backend?** One option is to dump the plugin's data collection output into a local file inside the plugin container. This is then read, parsed and format-validated [99] by the software core, then emitted to the monitoring backend. Or, the plugin container may be allowed to emit collected data directly to the backend over a secure communication channel (with corresponding exception added into the iptables). Output format verification and rate throttling then becomes the backend's responsibility, although some throttling can be employed at host.

## 6 Evaluation

Here we evaluate the security posture of our sandbox, as well as it's applicability and overhead to systems data extraction.

### 6.1 Security Analysis

**Selection of Exploits.** In order to test the efficacy of our sandbox, we considered a comprehensive set of attack categories, and **verified the inability of the exploits to impact** the guest container or the host system. We focused on the categories from the Exploit Database–a public archive of exploits used by penetration testers [90]. These include: local & privilege escalation, denial of service (DoS), remote exploits, as well as web application exploits. We also considered all of the attack categories from the popular Hansman and Hunt's attack taxonomy [62], namely: virus, worms, trojans, buffer overflows, DoS, network attacks, password attacks, information gathering, information corruption, information disclosure, service theft and subversion attacks, combined across all of the attack classification dimensions of the taxonomy.

The first three columns of Table 2 show how the attack vectors we consider (column 1), to portray possible avenues of attack specific to a cloud monitoring setting and *covering each of the above categories*, map to them (column 2,3). An attack may map to several categories; only the closest matches are indicated in the Table, restricted by space. Actual reported vulnerbilities in data collection plugins do indeed fall under the attack vectors coverage, such as arbitrary command execution [6] and DoS [1], as do attacks against the software core such as privilege escalation [44].

| Exploits / attack vectors | Categories from Hansman Hunt Taxonomy | Exploit DB | Direct guest infection ability when run inside Host or Guest | Direct guest infection ability when run inside Sand-box | Neutralized how? | Examples |
|---|---|---|---|---|---|---|
| [OOS] Output falsification | Service Theft | - | N | N | Doesn't directly hurt the guest, indirect impact possible | Fake OOM signal → Policy alarm → guest re-instantiation or quarantine |
| [OOS] Output format exploitation | Buffer Ovrflow | - | N | N | Doesn't hurt the guest, indirect impact possible, preventable via interface audit | Buffer overflow → subvert software core → [privilege escalation] → destroy guest |
| [OOS] Compromised monitoring backend | Info Leak | - | N | N | Doesn't directly hurt the guest, indirect impact possible | Leaked ssh keys → remote shell |
| [OOS] Compromised host | Info Leak | - | N | N | -- | Handover guest secrets to entity with external N/W privileges |
| [OOS] Kernel bugs | Subvert; DoS | PrivEsc, DoS | Y | Y | -- | Dirty Cow CVE-2016-5195: gist.github.com/rverton, Metasploit: BPF Priv Esc CVE-2016-4557 |
| [OOS] Insecure setuid binaries in guest | Subvert | PrivEsc | Y | Y | -- | Local privilege escalation → shell, Metasploit: HP smhstart OSVDB-91990 |
| [OOS] Weakly configur--ed applications | Passwd; N/W | Web App | Y | Y | -- | Local (unix-socket-based) MySQL access via weak password |
| Kernel-level rootkits | Trojan; Subvert | PrivEsc | Y (N in guest) | N | Linux capability not given to sandbox or guest | Adore (syscall hooking), SucKIT (/dev/kmem) |
| Data corruption / tampering | DoS Virus Corrupt | DoS | Y | N | Guest rootfs mounted as read-only (RO rootfs) | Ransomware |
| Fork bomb, Memory hog, CPU hog | DoS | DoS | Y | N | Separate cgroups | [shell] :(){ :\|: & };: [shell] stress -c[-m] 1 |
| Port hijacking | DoS | DoS | Y | N | bind() limited via selinux / disabled via seccomp | Prevent webserver init by hoarding ports 80,8080 |
| Zip bomb | DoS Virus Trojan | DoS | Y | N | RO rootfs, separate cgroups | Metasploit: gzip memory bomb DoS . Evilarc: github.com/ptoomey3/evilarc |
| Code injection, process corruption/termination | Subvert; Corrupt | PrivEsc, DoS | Y | N | Privilege separation, blocked ptrace(), RO /proc/<pid>/mem, no /dev/mem | sigsleeper: github.com/cys3c/PELT , http://pyrasite.com/ |
| Library hooking (LD_PRELOAD) | Subvert | PrivEsc | Y | N | Can't change guest proc (no ptrace()), RO rootfs, preloading doesn't work with suid binaries | vlany: github.com/mempodippy/vlany |
| Covert execution | Service Theft | - | Y | N | Any procs (hidden or not) in sandbox harmless to guest, visible on host, separate cgroups | mimic: github.com/cys3c/PELT , processhider: github.com/gianlucaborello/libprocesshider |
| Backdoor, reverse shell, botnet | Worm; N/W | Remote | Y | N | No N/W communication with outside world | vlany: github.com/mempodippy/vlany |
| Privilege escalation (remote & local via lo) | Subvert; N/W | PrivEsc, Remote | Y | N | No N/W communication with outside world, lo interface communication also blocked in netfilter | Metasploit: Samba cmd exec CVE-2007-2447 , Metasploit: java deserialization CVE-2008-5353 |
| Local privilege escalat--ion via filesystem | Subvert | PrivEsc | Y | N | RO rootfs, blocked lo interface communication | Metaploit: chkrootkit local privilege escalation CVE-2014-0476 |

Table 2: Exploit exploration summary: **Our sandbox is able to thwart all in-scope exploits ('N' in Column 5).** Abbreviations: 'OOS': Out-of-scope, 'Info Leak': information gathering and disclosure, 'Corrupt': information corruption, 'N/W': network attack. 'lo': loopback

While we test our sandbox against specific instance(s) of each type of exploit, we believe it to be generally applicable against other implementations of the exploits as well, since the **sandbox goes after plugging the fundamental avenues of attack, rather than any instance-specific vulnerability.** To this end, we have also opensourced our sandbox [24] so that it may be openly tested. Later in this Section, we highlight the specific design choices which neutralize each attack category.

**Execution of Exploits.** The last column of Table 2 mentions examples and sources of the particular exploit instances we tested in our sandbox. In some cases, we ran the exploits using the Metasploit penetration testing framework [13] running inside the plugin container, while the guest container runs a highly vulnerable Metasploitable image [14]. The corresponding metasploit module (containing the exploit code) then acts as an untrusted plugin, trying to subvert the target guest container it runs against. While in other cases, we got the exploit code from github, or implemented our own. This was then run directly inside the plugin container, as malicious actions of a hypothetical plugin.

**Direct vs. Indirect Impact.** Some of the exploits inherently cannot directly hurt the guest, although they can cause an indirect impact. For example, a code injection attack has a direct process-corruption impact, but not an output falsification attack, where the plugin falsely states that the guest is out of memory, potentially triggering a policy-driven guest re-instantiation. This is indicated as a 'Y' or an 'N' in columns 4 and 5, referring to the ability, or lack thereof, of an exploit to directly infect the guest, when run inside the guest or the host (column 4), and the sandbox (column 5).

**Impact of Out-of-Scope Exploits.** The first 7 attacks are out-of-scope of our threat model (Section 3.1) and are highlighted as 'OOS' in Column 1 of Table 2. The first 4 of these are indirect-impact causing, like the aforementioned output falsification attack. Another indirect-impact case is that of a compromised monitoring backend, where, for example, the adversary gains access to guest's credentials shipped by the plugin, enabling remote access to the guest. Similar is the case of a compromised host, where the plugin is able to leak sensitive guest information to a host-local accomplice via host-level side channels. A fourth indirect-impact attack vector is output format exploitation, where a malicious plugin writes badly formatted data to its output file, in the hope of exploiting programming errors in the software core (e.g. buffer overflow). A successful subsequent subversion or privilege escalation can then potentially impact the guest negatively. For that reason, it is important to carefully audit the interface to the core, which is unique to each software [97]. Output volume-based exploits, on the other hand, are in-scope and isolated using the blkio (block IO) cgroup controller.

The rest of the out-of-scope exploits are powerful enough to cause a direct guest impact. This includes attacks against a weakly configured guest. The sandbox cannot guard against the case, when, for example, the guest has insecure setuid binaries [80] lying around. Since the plugin has access to the read-only mounted guest rootfs, it can execute such a binary[5] to potentially escalate is privileges to that of the guest user, enabling full write-access to all of guest state. Another instance is that of a weakly configured guest application. If such an application is accessible over local/unix sockets with permissive DAC controls, the plugin can use password attacks to authorize itself to modify application state.

A buggy kernel is also out-of-scope, although, in some cases, running kernel-bug-exploitation based plugins within the sandbox may reduce their impact, as opposed to when they're run on the host or the guest. Consider for example the Dirty Cow privilege escalation exploit [46]. It exploits a race condition bug in the kernel code, where an unprivileged user (the plugin) can gain write access to otherwise read-only memory mappings and thus increase its privilege. On being executed, the exploit overwrote the plugin's read-only memory mappings of the guest's /usr/bin/passwd executable, and injected a shell payload inside it, subsequently enabling a shell access to the guest with the guest user's privileges. However, even after a successful exploit, although the sandboxed plugin could kill guest processes, but it was unable to modify guest's rootfs (read-only) or steal guest's secrets (networking disabled), unlike if it would have been guest or host-resident.

**Impact of In-Scope Exploits.** The different constraints added to the sandbox enable it to **contain all of the in-scope exploits** (row number 8 onwards), which would have otherwise directly impacted a guest container, assuming they were shipped inside third-party plugins. This is indicated as an 'N' in Column 5, referring to the inability of an exploit to infect the guest, when the former is run inside the sandbox. Compare this to the scenario when the exploit (masquerading as a legitimate plugin) is run inside the guest or the host–a 'Y' in Column 4, indication successful guest infection.

Column 6 shows how the different exploits get thwarted inside our sandbox. Kernel-level rootkits are trivial to guard against as the corresponding Linux capabilities to install kernel modules or access /dev/kmem are not given by default to unprivileged containers, including guests. Data corruption attacks, which can cause DoS to the guest by blocking its access to its own files (for example, encryption by a ransomware), are also easy to prevent by virtue of only read access provided to the plugin sandbox over guest's files, as opposed to write.

The next set of attacks also cause DoS by exhausting process counts, or hogging CPU, memory, disk or network resources. We tested these attacks by running inside the sandbox: (i) a trivial shell fork bomb, (ii) the stress resource hog utility, (iii) a zip bomb that explodes in space during unzipping, and (iv) a port hijacking script that tries to hoard all ports for itself so that the guest processes have none left to bind to. We verified containment of these DoS attacks, as well as continued access to its allocated resources by the guest. These operations of a hypothetical plugin are unable to impact the guest, due to their isolation inside a separate cgroup than the guest, and resource-limit enforcement. The attack-specific details are shown in rows 10-12 of Table 2.

The next set of exploits try to alter the guest's memory state, by corrupting or crashing any guest process, injecting malicious code inside it, or duping it to link to malicious libraries. Direct signaling (such as to suspend or terminate a process) is thwarted via privilege separation[6]. Most of these attacks require either (i) ptrace() to attach to the target process (guest's), which is however blocked for the sandbox, or (ii) write access to the guest process' /proc/pid/mem, which is only available as read-only to the plugin, owing to DAC controls, or (iii) /dev/mem-based access to global system memory, which is not mounted inside the unprivileged plugin sandbox. Furthermore, escalation may not be possible with library hooking, which does not work with setuid binaries. Then, without direct process modification, library hooking requires a guest-initiated process be linked to a malicious library which, however, does not exist on the guest's rootfs, while the latter also being unmodifiable (mounted read-only). The last column of Table 2, in the 5th-and-6th-last rows, lists the specific instances of these categories of attacks, which we verified with our sandbox.

---

[5]Restrictive DAC controls can help mitigate this.

[6]Permission checks for sending signals can be bypassed via CAP_KILL capability, which is not granted to the plugin.

With direct process modification disabled, another nefarious avenue is covert operations via backdoors, reverse shells, processes with fake names, for-profit botnets, and the like. But due to their reliance on network access, this is neutralized by cutting off the sandbox' communication with the outside world. Furthermore, hidden processes, or processes under fake names, still remain deprivileged and under all of the sandbox-enforced constraints, making them harmless to the guest, visible on the host [57], and impact-guarded via separate cgroups.

This network-access block, including via the loopback interface, also averts privilege escalation attacks against vulnerable applications, which may already be running inside the guest container. We verified this by running metasploit exploits from within the sandbox against multiple guest applications. One technique is to exploit via crafted inputs, as is the case with exploits targetting a stack buffer overflow vulnerability in MySQL (CVE-2008-0226), and a command execution vulnerability in Samba (CVE-2007-2447), for example. In the latter case, a shell can be obtained by the plugin under the guest's credentials, by connecting to Samba using a username containing shell meta characters (using its username map script feature). No authentication is needed to exploit this vulnerability since the username mapping happens prior to authentication. Another technique the plugin can employ, is to set up a malicious webserver for the guest container to inadvertently connect to. The de-serialization vulnerability in JVM (CVE-2008-5353) is one such example, in which case the java applet that subsequently runs, exploits a flaw in the JVM's deserialization of Calendar object on the guest side, to execute a shell for the plugin under the guest's privileges. The sandbox was able to successfully thwart these attacks against guest applications, which otherwise succeeded when the unprivilged plugin ran on the host or the guest, as shown in the 2nd-last row of Table 2.

With the network route blocked, a disk-based route is also available for privilege escalation attacks by the plugin, owing to its host-local proximity to the guest container. An interesting case is a vulnerability exploitation in the `chkroot-kit` [34] anti-rootkit software itself! In this case (CVE-2014-0476), a plugin can elevate its privileges simply by creating a file named `update`, with a shell payload, inside the guest's or host's `/tmp/` directory. Such disk-based attacks however trivially fail in case of the sandbox, since the guest's rootfs is accessible as read-only.

**Summary:** *By blocking the potential avenues of attack, our sandbox comprehensively protects the guest from malicious plugin behaviour, while allowing legitimate system-state-extraction operations.*

### 6.1.1 Discussion
(i) **chroot()** is not a security hole for us; we employ it only for existing libs to operate in an expected in-guest view. The actual protection is from the stronger *namespace* abstraction.

(ii) **Base container integrity.** We also consider attacks which attempt to break out of the base namespace abstraction / docker containerization abstraction as being out-of-scope, and classify them under the 'kernel bugs' category. As the containerization abstraction matures, these 'attacks' will eventually be neutralized, e.g. Shocker [53]. There exist complementary work which target containerization security [54, 103].

(iii) **Privacy.** Although a compromised monitoring backend is out-of-scope of our threat model, we can still help mitigate corresponding exploits. We can employ network sniffers or look at the plugin's output contents to search for 'sensitive information', which can be blocked from being shipped out to the backend. We can also guard against encrypted plugin output, by pro-actively blocking reads to sensitive files/dirs.

## 6.2 Applicability
The creation of the sandbox itself is independent of a data collection tool, and follows the implementation laid out in Section 5 To use this sandbox, however, the tool needs to be modified to support a 'sandbox mode'. This mode calls the sandbox creation routine, and implements the command-and-control interaction hooks with the sandboxed plugins, as discussed in Section 5.1, which are tool-dependent. To test the applicability of our sandbox, we selected an extensible opensource tool- ASC [36], because it already supports modular runtime modes—in terms of state extraction from VMs, containers, and hosts. Adding the sandbox mode to it boiled down to inheriting a base data-collector class and extending it to run ASC's plugins inside our sandbox, instead of running them on the host or the guest container.

We were able to run all of its plugins which are relevant in the setting of containerized guests running on a host system, except for application-level plugins such as Apache and Redis (coverage: 15/19 plugins). This is under the sandbox's strict security-over-visibility restriction which blocks querying local services (e.g. apache status webpage). For ASC, relaxing this restriction to allow HTTP GETs over (guest's) localhost, enables the sandbox to support all of ASC's plugins, without compromising on security, as discussed in Section 4.1.

For usability, it is important for the sandbox to be able to run existing plugins with no or minimal modifications, so as to be able to reuse existing plugin code. For ASC, most plugins ran unmodified. While in other cases, only minor modifications were needed- (i) avoiding `setns()` namespace jumping [21] for PID and network namespace, since the plugins already share them with the guest container, (ii) calling `chroot()` over the guest's R/O-mounted rootfs, to reuse existing disk-based data collection logic, and (iii) directing the CPU and memory plugins to gather guest's resource usage stats from the guest's cgroup filesystem mounted locally in the sandbox.

The outputs from the plugins were verified to match when run within and without the sandbox mode.

We analyzed the sandbox' applicability to Nagios and Sensu tools as well, which have many more plugins owing to their maturity and popularity in the community. We did not modify these tools to support the sandbox mode, but analyzed their plugins' code to ascertain possible execution under sandboxing. Amongst the official Nagios plugins, 42 are applicable to a cloud environment. Amongst those, the sandbox can support all 28 plugins which collect system-local information. The rest 14 talk to auxiliary system services over the network, such as DNS, SMTP, DHCP, NTP, ping and the like, which is disabled for the sandbox.

There are no 'official' Sensu plugins, so we analyzed its community plugins from its Github organization. Similar to the Nagios analysis, the sandbox can run 71 of the 80 system-level checks (`sensu-plugins-*-checks` repositories), the unsupported ones being the hardware-level checks interfacing with RAID controllers, and hardware sensors. Allowing read-only localhost access (Section 4.1), allows the sandbox to also support the application-level plugins as well, such as those collecting stats for Cassandra, Mongodb, Apache, etc. However, plugins which call service APIs over the network to collect data, such as from Amazon Web Services, Kubernetes, and Jenkins, etc., suffer under sandboxing, which doesn't allow communication with outside world.

**Takeaways:** *(i) The applicability coverage for system-level data extraction plugins is high. (ii) Since a majority of application-level plugins across ASC, Sensu and Nagios community exchange [87], follow the same approach to query application status, plugin-applicability coverage can be vastly improved by relaxing HTTP GETs over localhost (guest's; namespaced), without compromising on security. (iii) Coverage can potentially be further improved, if a trusted* `host:port` *whitelisting scheme can be enabled for plugins which query services over the network. (iv) 'Active events vs. isolation tradeoff' (Section 4.1) is not so pronounced, as many of the sandbox-able plugins already include active events surfaced via* `procfs`.

### 6.3 Performance

In this set of experiments, we measure the overhead our sandbox introduces to a normal systems data collection flow, in terms of sandbox creation latency, running time degradation, and resource consumption. *Setup*: We ran our experiments on a Ubuntu 16.04 KVM VM, with 4 vCPUs and 8G RAM, running Docker 1.12.1. The VM runs on quad-core / 16G RAM / Intel Core-i7 2.80GHz host machine, running Ubuntu 16.04 and QEMU 2.5.0. We use `python:2.7` as our Docker container image, to which we added all of ASC's plugins. All reported results are averages over 10 runs.

First off, it takes 344ms to setup the sandbox, compared to 273ms to run a guest container. Next, to measure the overhead of running the plugins inside our sandbox, we compared ASC's data collection cycle duration while running it's plugins in two configurations- (i) inside a guest container, and (ii) inside our plugin container prototype. We selected all of the plugins (as in Section 6.2) to run in each data collection cycle of ASC. A small difference was recorded in the running time of the plugins- none in the initial cycle (210ms), but an increase from 26ms/cycle (inside guest) to 30ms/cycle (inside sandbox), when amortized across 300 continuous cycles.

Finally, no extra resource consumption overhead was observed when running the plugins inside the sandbox as opposed to the guest container. The base memory usage of an inactive sandbox container is 176KB (same as any regular container), which rises to 19.6MB after 300 continuous data collection cycles- precisely the memory used by the guest container if the plugins are run inside it instead. The plugin container's CPU usage is also similar to the case of the guest container running the plugins.

We also verified that no extra sandboxing-related overhead is imposed upon the guest container workload. Although non-stop data collection cycles impact a sysbench CPU benchmark running inside the guest by 3.2%, the impact is same irrespective of where the plugins are run- on the host, the sandbox or guest itself (separate core). It is as per expectations that no extra impact be observed, owing to cgroups-based isolation.

## 7 Related Work

Existing approaches which may be employed for plugin verification or sandboxing can be categorized as follows:

**Privilege separation:** Compartmentalization is one of the first building blocks towards application security, and can be done manually via application restructuring [65, 84, 97], but with significant programmer effort [115], or automatically via program analysis [28, 30, 61]. Fortunately, most extensible software have this logical partitioning–core vs. plugins. We employ privilege separation in our sandbox, by running the plugins as unprivileged entities, with access to privileged guest state being mediated by other kernel constructs.

**Code analysis:** One way to validate a third-party plugin code is to employ code analysis techniques [26, 40, 45, 48, 50, 71, 111], which may help detect and avoid attacks such as buffer overflows, format string vulnerabilities, and API misusages. However, instead of exploiting programming errors, the plugin may employ other mechanisms, such as code obfuscation [37, 73], to perform nefarious actions (e.g., DoS, leaking secrets, or acting as botnet). Our sandbox, on the other hand, contains the exposure while remaining agnostic of any plugin code.

**Code transformation:** Another approach can be to transform the plugin code so that its instructions can be verified before they are allowed to execute, so as to satisfy any security policy. This transformation can be done statically (SFI) [49, 112, 117], or dynamically (SDT) [52, 104]. However, these high complexity solutions can impose a significant overhead on program execution, are limited to specific architectures, and provide less assurance than simpler hardware-based mechanism [56]. Our sandbox, on the other hand, uses generic OS-exported functionality, and has a low impact on plugin execution.

**System Call Interposition (SCI):** Syscall-based filtering [22, 27, 33, 41, 56, 58, 63, 67, 69, 82, 96] can also be employed to allow or deny a plugin's access to privileged/sensitive system state. Given the scale (and constant growth) of syscalls in the Linux kernel [77], creating and maintaining a robust-yet-generic syscall filter policy becomes a very complex and error-prone task [115]. Some SCI-based tools can also be circumvented, e.g. via 'time of check/time of use' race condition exploitation [55, 114]. Also, a syscall access policy alone does not suffice, since it might be necessary to lock down the system view visible to a plugin (Figure 2). That's why our sandbox employs other building blocks in addition to this.

**Language-based sandboxing** A programming language runtime barrier can also be used to restrict a plugin's operations and access to privileged state. This is usually accompanied by a restriction in the language functionality, such as limiting the set of external modules which can be loaded, or whitelisting and making internal built-ins read-only [23, 31, 32, 60, 64, 76, 79, 81, 113]. However, this turns out be a complex and error-prone task [93]. For example, an author's attempt to provide language-level sandboxing purely in python failed, to an extent where it was deemed necessary to put the whole Python process inside an external sandbox to guarantee security [110]. Furthermore, any bug in the programming language VM can still pose a threat [92]. Consequently, alternative language sandboxing solutions resort to OS-level sandboxing approaches [47, 59, 89, 98]. Our sandbox also uses OS-level constructs, thereby sidestepping the pitfalls of enforcing such language restrictions.

**OS isolation:** The kernel constructs to manage access rights and restrictions, by themselves, are insufficient for comprehensive sandboxing. As pointed out in [115], DAC/MAC are inadequate for application privilege separation. Fine-grained Type Enforcement policies (as in SElinux) are inflexible, difficult to write and maintain, and thus, in practice, broad rights are conferred. Chroot limits only file system access, and switching credentials via setuid offers poor protection against weak DAC protections on namespaces. Namespaces-based view separation itself precludes cross-domain (e.g., a container) visibility. Linux capabilities, in their current form, still confer too much power than required for fine-grained access control. Capsicum [115] enables finer granularity capabilities via file-descriptor-level access control. However, since it combines security policy with code in the application, this makes it harder to cleanly specify and analyze a security policy. It also requires kernel modifications, and (minor) application modifications to make use of the proposed kernel construct.

As we've shown, a combination of these constructs is needed to get the right mix of accessibility and restrictions required for sandboxing systems data collection plugin.

**Hardware virtualization:** Hardware virtualization primitives can also be used to isolate plugin code. Examples of application-level sandboxing include KVMsandbox [25] and libvirt-sandbox [42]. Microvirtualization (Bromium [29]), compartmentalization (QubesOS [100]), and unikernel [75] approaches are also potential sandboxing options. Such a setup would still require secure mechanism to provide VM-to-VM or VM-to-host visibility (required for state extraction plugins). We are able to leverage kernel-exported functionality to easily achieve this in our sandbox design.

Amongst the security solutions employing a combination of techniques for different use-cases [51, 68, 72, 78, 102], the Firejail [51] sandbox comes closest to our approach. Although its set of kernel control knobs is similar to ours, but the manner of tuning those knobs differs since it needs a different blend of access and control owing to its different target usecase. Taking example of just namespaces alone, in case of firejail each sandboxed application is give its own private set of namespaces. This would not work for the sandboxed plugins of a monitoring software, since they need to access the target endpoints' namespace to extract relevant system state. Our sandbox facilitates this access in a secure manner.

WatchIT [106] also uses containers to create a sandbox. But its usecase, as well as the capabilities and formation of its sandbox is very different to ours. Taking just one example of its sandbox' permissions, since it allows (controlled) root access it becomes too permissive for data collection plugins. But this makes sense for the use-case it targets–IT admins and third-party contractors, who may need root access to perform necessary IT actions. Our sandbox is able to provide a read-only, non-root access sufficient for third-party plugins. Furthermore, since WatchIT's sandbox is much more permissive than ours, it requires potentially heavyweight monitoring and logging of network traffic and filesystem accesses, to avoid potential attacks, which we don't. WatchIT also requires kernel modifications unlike our solution.

## 8 Conclusion

In this paper, we have presented our sandbox mechanism to enable secure extensibility for systems data extraction software. We described how we protect the software core and the monitored guest from potentially malicious plugins, by isolating the plugins inside a sandbox environment, while allowing legitimate plugins to collect system state, by granting them read-only visibility into the guest system. We presented a survey of existing monitoring software to highlight the need for a secure plugin sandbox. We highlighted the strong security posture of our sandbox, by verifying successful containment of several exploits across multiple dimensions. We demonstrated its applicability and low state extraction overhead, by sandboxing plugins of an existing data collection software against containerized guest systems. We have opensourced our sandbox, and invite community feedback.